\documentclass[%
reprint,
superscriptaddress,
nofootinbib,
nobibnotes,
amsmath,amssymb,
aps,
]{revtex4-2}
\usepackage[latin1]{inputenc}
\usepackage{graphicx}
\usepackage{dcolumn}
\usepackage{bm}
\usepackage[hidelinks]{hyperref}
\usepackage{multirow}
\usepackage{xcolor}
\usepackage{braket}

\begin{document}

\title{Lifetime of the $^2F_{7/2}$ level in Yb$^+$ for spontaneous emission of electric octupole radiation}
\author{R.~Lange}
\affiliation{Physikalisch-Technische Bundesanstalt, Bundesallee 100, 38116 Braunschweig, Germany}
\author{A.~A.~Peshkov}
\affiliation{Physikalisch-Technische Bundesanstalt, Bundesallee 100, 38116 Braunschweig, Germany}
\affiliation{Institut f\"ur Mathematische Physik, Technische Universit\"at Braunschweig, Mendelssohnstra\ss{}e 3, 38106 Braunschweig, Germany}
\author{N.~Huntemann}
\email{nils.huntemann@ptb.de}
\affiliation{Physikalisch-Technische Bundesanstalt, Bundesallee 100, 38116 Braunschweig, Germany}
\author{Chr.~Tamm}
\affiliation{Physikalisch-Technische Bundesanstalt, Bundesallee 100, 38116 Braunschweig, Germany}
\author{A.~Surzhykov}
\affiliation{Physikalisch-Technische Bundesanstalt, Bundesallee 100, 38116 Braunschweig, Germany}
\affiliation{Institut f\"ur Mathematische Physik, Technische Universit\"at Braunschweig, Mendelssohnstra\ss{}e 3, 38106 Braunschweig, Germany}
\affiliation{Laboratory for Emerging Nanometrology, Langer Kamp 6a/b, 38106 Braunschweig, Germany}
\author{E.~Peik}
\affiliation{Physikalisch-Technische Bundesanstalt, Bundesallee 100, 38116 Braunschweig, Germany}

\date{\today}

\begin{abstract}
We report a measurement of the radiative lifetime of the $^2F_{7/2}$ level of $^{171}$Yb$^+$ that is coupled to the $^2S_{1/2}$ ground state via an electric octupole transition.
The radiative lifetime is determined to be $9.96(50)\times 10^7$~s, corresponding to 3.16(16) years. 
The result reduces the relative uncertainty in this exceptionally long excited state lifetime by one order of magnitude with respect to previous experimental estimates.
Our method is based on the coherent excitation of the corresponding transition and avoids limitations through competing decay processes. 
The explicit dependence on the laser intensity is eliminated by simultaneously measuring the resonant Rabi frequency and the induced quadratic Stark shift. 
Combining the result with information on the dynamic differential polarizability permits a calculation of the transition matrix element to infer the radiative lifetime. 
\end{abstract}

\maketitle

\textbf{Note: In an earlier version of this paper, we erroneously used the electric field amplitude $E_0^2$ of the perturbing laser radiation instead of the mean-square value of the alternating electric field $\langle E^2\rangle = E_0^2 /2$ in the quadratic Stark shift $\Delta \nu_{\text{QS}}$. The measured differential polarizabilities, however, were based on the quadratic Stark shift now given in Eq.~(1). Consequently, the lifetime was underestimated by a factor factor two and the matrix element overestimated by $\sqrt{2}$. This has been corrected in the current version.}

Coherent interrogation of trapped particles facilitates the determination of atomic transition frequencies with high accuracy, recently demonstrated below the 10$^{-18}$ fractional uncertainty level~\cite{bre19}. 
Complementary information on the electronic structure of atomic systems can be obtained from measurements of coupling strengths between electronic states and of their natural lifetimes. 
For transitions with coupling via optical electric dipole (E1) radiation, excited states show natural lifetimes of nanoseconds that can be determined directly with a relative uncertainty at the $10^{-3}$ level by observing the spontaneous decay (see for example~\cite{olm09,bel12}).
Even higher precision has been obtained by using theoretical information on the atomic structure together with measurements of dynamic Stark shifts resulting from the E1 coupling to Rydberg states~\cite{woo10}. 
As recently demonstrated, by combining measurements of transition rates and Stark shifts from the same laser, an explicit dependence on its intensity is avoided and the E1 coupling strengths can be determined precisely without theoretical modeling~\cite{het15}.
Low-lying electronic states for which decay is not supported by E1 selection rules show significantly longer natural lifetimes. 
Such meta-stable states of isolated quantum systems provide the basis for application in frequency metrology, quantum simulation, and quantum information processing~\cite{lud15,bla12}. 

If the natural lifetime exceeds seconds, a direct observation of the spontaneous decay is particularly challenging because of competing processes such as collisions with the background gas, off-resonant laser radiation, and the long measurement periods needed to achieve sufficiently small statistical uncertainty~\cite{bar00,ros07,sha16}.
Using large ensembles of laser-cooled neutral atoms confined in magnetic quadrupole traps, it has been possible to investigate excited state lifetimes exceeding minutes~\cite{yas04,jen11}. 
With this technique, a lifetime $\tau = 7870(510)$~s has been found for the $2^{3}S_1$ state of Helium, to our knowledge the longest natural lifetime on an optical transition determined experimentally so far \cite{hod09}.
For even longer lifetimes, the competing processes dominate and make a direct measurement of the natural decay time increasingly difficult. 
\begin{figure}[b]
{\centering \includegraphics[width=.95\columnwidth]{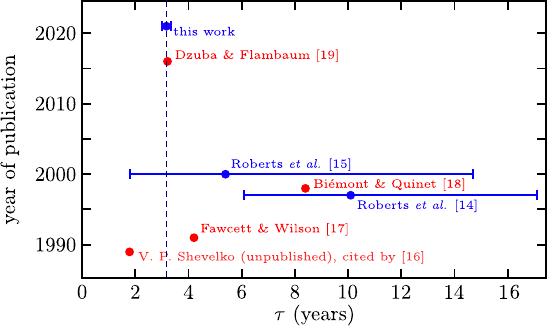}}
\caption{Comparison of experimental (blue) and theoretical (red) values of the Yb$^+$ excited $^2F_{7/2}$ state lifetime $\tau$. Previous experimental estimates \cite{rob97,rob00} were based on observed laser excitation events and a  rate equation analysis. Theoretical values are given in Refs.~\cite{leh89,faw91,bie98,dzu16a}.}
\label{graph:Lifetime}
\end{figure}

In this Letter, we devise an alternative method to determine the natural lifetime of metastable states for systems with a single radiative decay channel from an excited state $\ket{e}$ to a ground state $\ket{g}$.
The method relies on  monitoring the coherent time evolution  of the two-level system  while it is resonantly excited by a laser, in order to extract the matrix element of the transition.
To illustrate the method, we investigate the ${}^2S_{1/2} (F=0) \rightarrow {}^2F_{7/2} (F=3)$ electric octupole (E3) transition in a single trapped ${}^{171}\text{Yb}^+$ ion and infer the yearslong natural lifetime of the excited ${}^2F_{7/2}$ state with 5\% uncertainty.
As shown in Fig.~\ref{graph:Lifetime}, our result reduces the relative uncertainty by one order of magnitude with respect to previous long-standing estimates~\cite{rob97,rob00} and provides a clear point of reference for atomic structure calculations.

The method presented here is based on the fact that the radiative lifetime $\tau$ is related to the matrix element $V_{eg}$ of the transition according to Fermi's golden rule, $\tau \propto 1/|V_{eg}|^2$, where the proportionality coefficient depends on the experimental geometry and the angular momentum of ground and excited state.
If the transition is coherently driven by a laser, the matrix element is measurable from the Rabi frequency $\Omega$ describing the oscillation of population between $\ket{g}$ and $\ket{e}$ states: $\Omega = 2\pi E_0 |V_{eg}|/h$, with $h$ the Planck constant.
Independence from the electric field amplitude $E_0$ of the laser at the ion position can be achieved by simultaneously measuring the differential quadratic Stark shift $\Delta \nu_{\text{QS}} = - \langle E^2\rangle \Delta \alpha_{eg}(\nu_0) / (2h)$ and defining the relative excitation strength $\xi = \Omega^2 / \Delta \nu_{\text{QS}}$.
This quantity together with the differential polarizability $\Delta \alpha_{eg}$ at the transition frequency $\nu_0$, obtained independently from experiments or theory, permits the determination of the matrix element:  $|V_{eg}| = 1/(2\pi) \sqrt{h |\xi \Delta \alpha_{eg} (\nu_0)|/4}$.

The Yb$^+$ E3 transition is employed in optical atomic clocks~\cite{god14,san19} and currently features the most accurately determined transition frequency~\cite{lan21}. 
It is well-suited for various searches for physics beyond the standard model~\cite{saf18a} and has been used to realize the most stringent limits for potential violations of local Lorentz invariance in the electron sector \cite{san19}. 
Because of the large sensitivity of the transition frequency to variations of the fine structure constant $\alpha$, repeated comparisons to other frequency references currently provide the most rigorous constraints on a temporal variation of $\alpha$ and a coupling of $\alpha$ to gravity~\cite{lan21}. 
Besides applications in other searches for \textit{new physics}~\cite{cou20,all21}, {Yb}$^+$ is employed in a number of quantum computing experiments~\cite{mon21,edm21,kau18}, which can take advantage of the long-living ${}^2F_{7/2}$ state~\cite{edm21,mcm20}.

\begin{figure}
{\centering \includegraphics[width=.95\columnwidth]{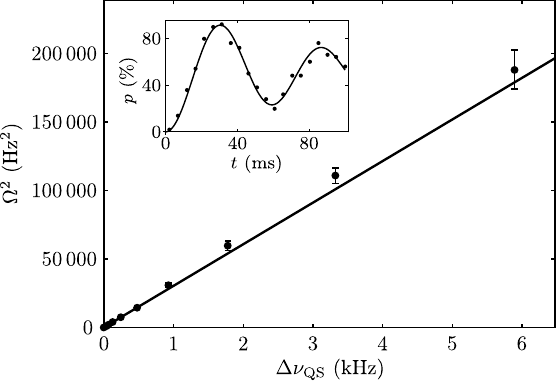}}
\caption{Investigation of the relative excitation strength  of the $^{171}$Yb$^+$ E3 transition. At different intensity settings of the probe laser, the quadratic Stark shift $\Delta \nu_{\text{QS}}$ and the resonant Rabi frequency $\Omega$ are measured. Each Rabi frequency is obtained from a Rabi oscillation as shown in the inset for $\Omega = 2\pi \times 19.2(3)$~Hz at $\Delta \nu_{\text{QS}} = 477(1)$~Hz, where $p$ is the $^{2}F_{7/2}$ state excitation probability and $t$ the Rabi pulse duration. Assuming a linear dependence between quadratic Stark shift and squared Rabi frequency yields the relative excitation strength $\xi = \Omega^2 / \Delta \nu_{\text{QS}} = 30.3(9)$~Hz.}
\label{graph:Lightshift}
\end{figure}

In our experiment, a single ion is confined in a radio frequency Paul trap at ultrahigh vacuum and laser-cooled on the ${}^2S_{1/2} \rightarrow {}^2P_{1/2}$ transition at 370~nm close to the Doppler limit.
During cooling, population trapping in the ${}^2D_{3/2}$ state is prevented using laser radiation at 935~nm.
Stray electric fields are compensated to suppress the relative strength of first-order micromotion sidebands to less than 1\%. 
The frequency of the 467~nm (642~THz) probe laser is stabilized via a frequency comb generator to the length of a cryogenic silicon cavity~\cite{mat17a} and permits coherent excitation of the E3 transition with laser pulses of up to 500~ms duration. 
After successful excitation, the ${}^2F_{7/2}$ state can be rapidly depopulated using laser radiation at 760~nm.
Without this repump laser, the ${}^2F_{7/2}$ state is quenched by collisions with the residual gas \cite{sei95,jau15b} and via excitation of the ${}^2F_{7/2}\rightarrow {}^2D_{5/2}$ transition by room-temperature thermal radiation, and a lifetime of several hours is typically observed.
The linear polarization $\bm{\epsilon}$ of the probe laser beam with wave vector $\bm{k}$ and the orientation of the ion quantization axis, defined by an externally applied magnetic field $\bm{B}$, are chosen to maximise the excitation probability: 
We align $\bm{\epsilon}$, $\bm{k}$ and $\bm{B}$ within one plane and set the angle $\beta_{\text{E3}}$ between $\bm{\epsilon}$ and $\bm{B}$ to $59(1)^{\circ}$~\cite{sch20e}.
To derive the resonant Rabi frequency, excitations of the E3 transition with variable pulse duration are performed.
Under ideal conditions with the ion in the motional ground state, the Rabi frequency can be directly deduced from the oscillation of the excitation probability, $p(t) = ( 1 - \cos(\Omega t))/2$.
However, in our experimental realization with a Lamb-Dicke parameter of about $0.08$, Doppler cooling to a mean motional quantum number of about $30$ leads to considerable damping of the oscillation due to the different couplings between ground and excited motional states~\cite{win98} as shown in the inset of Fig.~\ref{graph:Lightshift}.
Taking this into account, the Rabi frequency is determined for various settings of the probe laser intensity.
For each intensity the quadratic Stark shift is measured as the offset $\Delta \nu_{\text{QS}}$ from the unperturbed transition frequency $\nu_0$.
In order to determine $\nu_0$ and to correct for the frequency drift of the silicon reference cavity, the measurements are complemented by periods where the experiment is operated as an optical clock so that the laser frequency is locked to $\nu_0$~\cite{lan21}. 
Thereby, $\Delta \nu_{\text{QS}}$ is measured with a relative uncertainty of less than 1\%.
The results are depicted in Fig.~\ref{graph:Lightshift}. 
Assuming a linear dependence for the data, we obtain the relative excitation strength $\xi = 30.3(9)$~Hz. 
The uncertainty results from statistics, the residual motion of the ion and possible deviations from the resonant frequency during the determination of the Rabi frequency. 

In addition to the relative excitation strength for the E3 transition, the differential polarizability needs to be determined experimentally to infer the matrix element.
The polarizability predominantly results from nonresonant couplings to higher-lying electronic levels and comprises a scalar part $\Delta \alpha_S$ and a tensorial part $\Delta \alpha_T$. 
Stark shift and polarizability are related through~\cite{ita00}
\begin{equation}
    \Delta \nu_{\text{QS}} (\nu) = -\frac{\langle E^2\rangle}{h} \left[ \frac{\Delta \alpha_S(\nu)}{2} + \frac{\Delta \alpha_T(\nu)}{5} (3\cos^2 \beta - 1) \right] \, ,
    \label{eq:stark1}
\end{equation}
where $\beta$ is the angle between the external magnetic field orientation and the linear polarization of the laser radiation of frequency $\nu$.
We assume a linear dependence of the differential polarizability around $\nu = \nu_0$ since the resonant contribution of the E3 transition is negligible.
To determine the differential polarizability, the ion is exposed to nonresonant perturbing laser radiation at frequency close to $\nu_0$.
The perturbing beam is focused to an approximately Gaussian spot and the induced quadratic Stark shift is measured for multiple transverse displacements with respect to the ion.
Summing the shift for three mutually orthogonal directions of the magnetic field averages the angle-dependent part of Eq.~\eqref{eq:stark1} to zero~\cite{ita00}.
The scalar differential polarizability $\Delta \alpha_S$ is derived from the spatially integrated quadratic Stark shift profile and the optical power of the beam.
With the ion at beam center, the tensor polarizability $\Delta \alpha_T$ is inferred from variations of the quadratic Stark shift for different orientations of the magnetic field.
In this way, using a perturbing field at 467(1) nm wavelength the values of $\Delta \alpha_S(\nu_0)$ and $\Delta \alpha_T(\nu_0)$ are measured together with the polarizabilities at various near-infrared wavelengths~\cite{hun16}. We find

\begin{align*}
    \Delta \alpha_S(\nu_0) &= -2.67(11) \times 10^{-40} \text{Jm}^2/\text{V}^2 \, , \\
    \Delta \alpha_T(\nu_0) &= 4.1(2) \times 10^{-41} \text{Jm}^2/\text{V}^2 \, . 
\end{align*}
The uncertainties result predominantly from the optical power measurement and its stability (3\%), beam profile imperfections (1.5\%), and the deviation of the perturbing laser from the E3 probe laser wavelength of less than 2~nm (1.5\%).
As described above, the matrix element of the E3 transition of $^{171}$Yb$^+$ is determined from the relative excitation strength and the differential polarizability to be
\begin{align}
   |V_{eg}(\text{E3})| = 1.85(5) \times 10^{-37} \text{Jm/V} \, . \nonumber
\end{align}

In order to use this value for a determination of the natural lifetime $\tau$, we recall general expressions from atomic structure theory. 
The matrix element for the coupling of a photon to a many-electron atom can be expressed as
\begin{align}
\label{eq:matelem}
    V_{eg} = \frac{ec}{2\pi \nu_0} \Bigg \langle \gamma_{e} F_{e} M_{e} \Bigg| \sum_{q} \bm{\epsilon} \, \bm{\alpha}_{q} \,  e^{i\bm{k} \bm{r}_q} \Bigg| \gamma_{g} F_{g} M_{g} \Bigg \rangle  \, ,
\end{align}
where the sum runs over all electrons in the atom and $\bm{\alpha} = (\alpha_x,\alpha_y,\alpha_z)$ denotes the vector of Dirac matrices, $e$ the elementary charge, and $c$ the speed of light.
The atomic states are specified by the total angular momenta $F_{g,e}$ and their projections $M_{g,e}$, as well as the additional quantum numbers $\gamma_{g,e}$. 
To meet our experimental conditions, we assume linearly polarized light with wave vector $\bm{k}$ and polarization $\bm{\epsilon}$ aligned with the magnetic field in one plane.

In order to simplify Eq.~\eqref{eq:matelem}, we make use of the multipole decomposition of the electron-photon interaction operator~\cite{ros57} along with the Wigner-Eckart theorem. 
Considering transitions between electronic states with total electronic angular momenta $J_g$ and $J_e$, with the nuclear spin $I$ unchanged, we obtain
\begin{align}
	\label{eq:matelem_2}
	V_{eg} &= \frac{\sqrt{\pi}ec}{2\pi\nu_0} \, 
	 i^L (-1)^{J_e+I+F_g+L} \sqrt{(2L+1)(2F_g+1)} \, \notag \\ 
	&\times \left\{ \begin{array}{ccc}
		F_e & F_g & L \\
		J_g & J_e & I
	\end{array} \right\} \, \langle F_g M_g \, L M_e \!-\! M_g | F_e M_e \rangle \, \notag \\
	&\times \langle \gamma_e J_e || H_{\text{ph}} (pL) || \gamma_g J_g \rangle \sum_{\lambda = \pm 1} (i\lambda)^p \, d^{L}_{M_e \!-\! M_g, \, \lambda} (\theta)  \, ,
\end{align}
where $d^{L}_{M_e \!-\! M_g, \, \lambda} (\theta)$ is the Wigner $d$-function, $\theta = {(90^{\circ} - \beta)}$ the angle between the photon wave vector $\bm{k}$ and the quantization axis of the atom, and $\langle \gamma_e J_e || H_{\text{ph}} (pL) || \gamma_g J_g \rangle$ the reduced matrix element.
We have restricted ourselves here to a single leading multipole term $(pL)$, in which $p=0$ and $p=1$ refer to magnetic and electric $2^L$-pole radiation, respectively. 

From Eq.~\eqref{eq:matelem_2}, we obtain the rate for the spontaneous decay $| \gamma_{e} J_{e} \rangle \rightarrow | \gamma_{g} J_{g} \rangle$.
This rate gives the probability per unit time for emission of a photon with multipolarity $(pL)$ and is given by~\cite{joh07}
\begin{align}
\label{eq:EinsteinA}
    R =& \frac{16\pi^2 \alpha \nu_0}{2J_e+1} |\langle \gamma_e J_e || H_{\text{ph}} (pL) || \gamma_g J_g \rangle|^2 \, ,
\end{align}
with $\alpha$ the fine structure constant.
The natural lifetime $\tau$ of the excited state $| \gamma_{e} J_{e} \rangle$ is obtained by the inverse of $R$.

In our experiment, we investigate an electric octupole transition ($L=3$, $p=1$) with $J_g =1/2$, $F_g =0$, $I = 1/2$, $J_e =7/2$, $F_e = 3$, $M_e =0$ and set $\beta = \beta_{\text{E3}}$.
From $|V_{eg}(\text{E3})|$ and Eqs.~\eqref{eq:matelem}-\eqref{eq:EinsteinA}, the natural lifetime of the $^{2}F_{7/2}$ state of $^{171}$Yb$^{+}$ is determined as
\begin{equation*}
    \tau(^2F_{7/2}) = 9.96(50)\times 10^7~\text{s},
\end{equation*}
corresponding to 3.16(16)~\text{years}.

The obtained value for the lifetime is compared to previous measurements and calculations in Fig.~\ref{graph:Lifetime}. 
Our result reduces the relative uncertainty by about one order of magnitude with respect to previous long-standing estimates. It represents to our knowledge the longest experimentally determined natural lifetime of an electronic state to date and the first precise measurement of an E3 radiative lifetime.
The achieved relative uncertainty of 5\% is mainly limited by the residual temperature of the trapped ion in the measurement of the relative excitation strength and by the determination of the differential polarizability. 
While the former contribution can be easily reduced by resolved sideband cooling~\cite{die89}, superior accuracy in the differential polarizability is obtained for other ion species such as $^{88}$Sr$^+$~\cite{dub13}, $^{40}$Ca$^+$~\cite{hua19}, or $^{138}$Ba$^+$~\cite{bar19b}. 
Co-trapping $^{171}\text{Yb}^+$ with such an ancillary ion permits a transfer of the relative accuracy in the differential polarizability.

Our experimental result for the lifetime has been obtained with $^{171}$Yb$^+$ with nuclear spin $I=1/2$ while earlier work has also investigated $^{172}$Yb$^+$ with $I=0$~\cite{rob97} (see Fig.~\ref{graph:Lifetime}). 
For the  $^{173}$Yb$^+$ isotope with nuclear spin $I=5/2$ a shortening of the lifetime of some of the hyperfine levels of the $^2$F$_{7/2}$ state by more than 2 orders of magnitude has been predicted~\cite{dzu16a}. 
The effect is due to the admixture of the $^2$P$_{3/2}$ state that is induced by hyperfine interaction with the nuclear electric quadrupole moment.
The lifetime of the unperturbed E3 decay that we have determined here is expected to be valid for all stable isotopes of Yb with the exception of $^{173}$Yb$^+$ and provides the reference for a quantitative experimental study of the hyperfine quenching effect in the latter~\cite{rom21}.

The method of combining measurements of the resonant Rabi frequency and the quadratic Stark shift for the determination of a small transition matrix element is readily applicable to other atomic species featuring electronic states with natural lifetimes exceeding minutes and promises high accuracy due to the immunity to competing decay processes. 
Particularly for highly-forbidden transitions in $^{175}\text{Lu}^+$~\cite{arn16}, Pb$^{2+}$~\cite{bel20b}, alkaline-earth atoms~\cite{der01}, and highly-charged ions~\cite{bek19} that are of interest for optical clocks and searches for violations of fundamental symmetries, accurate assessments of the lifetime support a better understanding of the atomic structure.

We thank Burghard Lipphardt, Thomas Legero, Erik Benkler and Uwe Sterr for providing the ultrastable laser reference Si-2.
We thank M. Steinel for noticing an error in v1 of this report and bringing it to our attention.
This work has been supported by the Max-Planck-RIKEN-PTB-Center for Time, Constants and Fundamental Symmetries.
Furthermore, this work has been funded by the Deutsche Forschungsgemeinschaft (DFG, German Research Foundation) -
Project-ID 274200144 - SFB 1227 within project B02 and under Germany's Excellence Strategy - EXC-2123 QuantumFrontiers - 390837967.

\end{document}